\documentclass{PoS}
\usepackage[utf8]{inputenc}
\usepackage[T1]{fontenc}
\usepackage{url}
\title{SMPDF Web: a web-based application for specialized minimal
  parton distribution functions}

\ShortTitle{SMPDF Web}

\author{\speaker{Stefano Carrazza}\thanks{CERN-TH-2016-129}\\
        Theoretical Physics Department, CERN, Geneva Switzerland\\
        E-mail: \email{stefano.carrazza@cern.ch}}

\author{Zahari Kassabov\thanks{TIF-UNIMI-2016-5}\\
        Dipartimento di Fisica, Universit\`a di Torino and INFN, Sezione di Torino\\
		TIF Lab, Dipartimento di Fisica, Universit\`a di Milano\\
        E-mail: \email{kassabov@to.infn.it}}

\abstract{ We present SMPDF Web, a web interface for the construction
	of parton distribution functions (PDFs) with a
	minimal number of 
	error sets needed to represent the PDF uncertainty of specific
	processes (SMPDF). }

\FullConference{VII Workshop italiano sulla fisica pp a LHC\\
		16-18 Maggio 2016\\
		Pisa, Italy}

\begin{document}

\paragraph{The PDF4LHC15 recommendation and tools}

The accurate determination of the parton distribution functions 
of the proton is crucial for precise predictions at the Large Hadron
Collider (LHC). The PDF4LHC working group has been tasked with
producing a recommendation for a standard method of calculating
PDF$+\alpha_S$ uncertainties.

The most recent PDF4LHC recommendation~\cite{Butterworth:2015oua}
prescribes a combined PDF set (branded PDF4LHC15), composed of the
statistical combination~\cite{Forte:2010, Watt:2012tq}  of the individual PDF determinations from
three independent collaborations: NNPDF3.0~\cite{Ball:2014uwa},
CT14~\cite{Dulat:2015mca} and MMHT2014~\cite{Harland-Lang:2014zoa}.
These PDF sets satisfy a set of compatibility requirements: usage of
global datasets for the PDF determination, theoretical predictions and
parton evolution computed in the General Mass Variable Flavour Number
Scheme (GM-VFNS) and $\alpha_s$ set to the PDG
average~\cite{Agashe:2014kda}.  The recommended usage and
applicability are discussed in Ref.~\cite{Butterworth:2015oua}.

The recommendation is implemented by first constructing a combined prior PDF
set of Monte Carlo PDF \emph{replicas} from each of the three
collaborations: A probability distribution related to the PDF
uncertainty of a given hadronic level observable can be computed by
convolving the corresponding parton level quantity with each of the
replicas. Then any statistical quantity (such as the mean and the
standard deviation) can be obtained from the resulting distribution.
This becomes impractical when the number of Monte Carlo replicas
required to faithfully reproduce the prior distribution is high enough
that a significant computational effort is required to perform all the
convolutions. This is the case for the PDF4LHC prior which contains
a total of 900 replicas. Therefore several compressed PDF sets are
delivered together with the prior.  They are based on three different
algorithms: Compressed Monte Carlo
(CMC-PDF)~\cite{Carrazza:2015hva,Carrazza:2016sgh}, Monte Carlo to
Hessian, MC2H~\cite{Carrazza:2015aoa} and Meta-PDF~\cite{Gao:2013bia}.
The latest two also transform the Monte Carlo sample into a Hessian
set, which is more adequate for certain experimental analyses, and
more efficient in the circumstances where one can assume that the
prior set is Gaussian. The CMC methodology excels at reproducing the
non-Gaussianities of the prior (particularly important for searches).
A study about the accuracy of these methodologies for different
observables is performed in Ref.~\cite{Badger:2016bpw}. The final
result is made available in the LHAPDF format~\cite{Buckley:2014ana}.

\paragraph{Specialized minimal PDFs}

As a follow-up of the studies that led to the PDF4LHC15
recommendations, a Hessian reduction algorithm called Specialized
Minimal PDF (SMPDF) was proposed and implemented in
Ref.~\cite{Carrazza:2016htc}. It is a development upon the MC2H
methodology, where the PDF covariance matrix is reproduced by
selecting the largest eigenvectors through Principal Component
Analysis (PCA) and expressing them as a linear combination of the
input Monte Carlo replicas.
In the SMPDF methodology we focus on providing a representation of the
covariance matrix, which allows an accurate determination of the PDF
uncertainties for a specific set of processes, with a minimal number
of PDF error sets.  This is achieved through an iterative procedure, in
which one eigenvector (error set) is added at a time until the
standard deviation for each of the input processes is reproduced
better than a threshold selected by the user. Therefore the threshold
is an upper bound for the inaccuracy on the input observables.

Different processes can be combined
(either in the same SMPDF set or from independent ones) in such a way
that the PDF correlation between them is reproduced, and further
information can always be added. To this end, we provide explicitly
the linear transformation that converts the prior set into the
resulting SMPDF.

The eigenvectors are selected based on kinematic considerations
(specifically the correlation between the value of a given PDF flavour
and point in $x$ with the value of the observable), which ensures that
the methodology is robust upon variations in the cuts and generalizes
efficiently to similar processes. Therefore one
can reliably use SMPDFs to compute predictions of processes that were
not explicitly given as input to the algorithm (or were given with
different cuts) but hold a similar PDF dependence.

The SMPDF methodology has been explicitly validated for a number of
representative Standard Model processes of particular relevance at the
LHC (including Higgs, top and electroweak physics). In each case we
observe a large reduction in the number of error sets while keeping an
accuracy comparable to that of the MC2H reduction. For example, as
discussed in Ref.~\cite{Carrazza:2016htc}, it is possible to reproduce
the predictions of the PDF4LHC prior for the most relevant Higgs
production channels with 15 error sets (to be compared with the 900 of
the prior and the 100 of MC2H). If one is only interested in the gluon
fusion channel, then only 4 error sets suffice.

As shown in Ref.~\cite{Badger:2016bpw}, by selecting a general enough
set of input observables (constructing what we call Ladder SMPDF), one
can achieve the same accuracy as Meta PDF, for a large set of Standard
Model processes, with about half as many error members and with the
possibility to trade better accuracy for more error sets, by
decreasing the threshold parameter and/or increasing the number of
input processes.  This shows that SMPDFs can be advantageous in
situations where both computational performance and accuracy in the
computation of PDF uncertainties are needed.

\begin{figure}[t]
  \begin{centering}
    \includegraphics[scale=0.35]{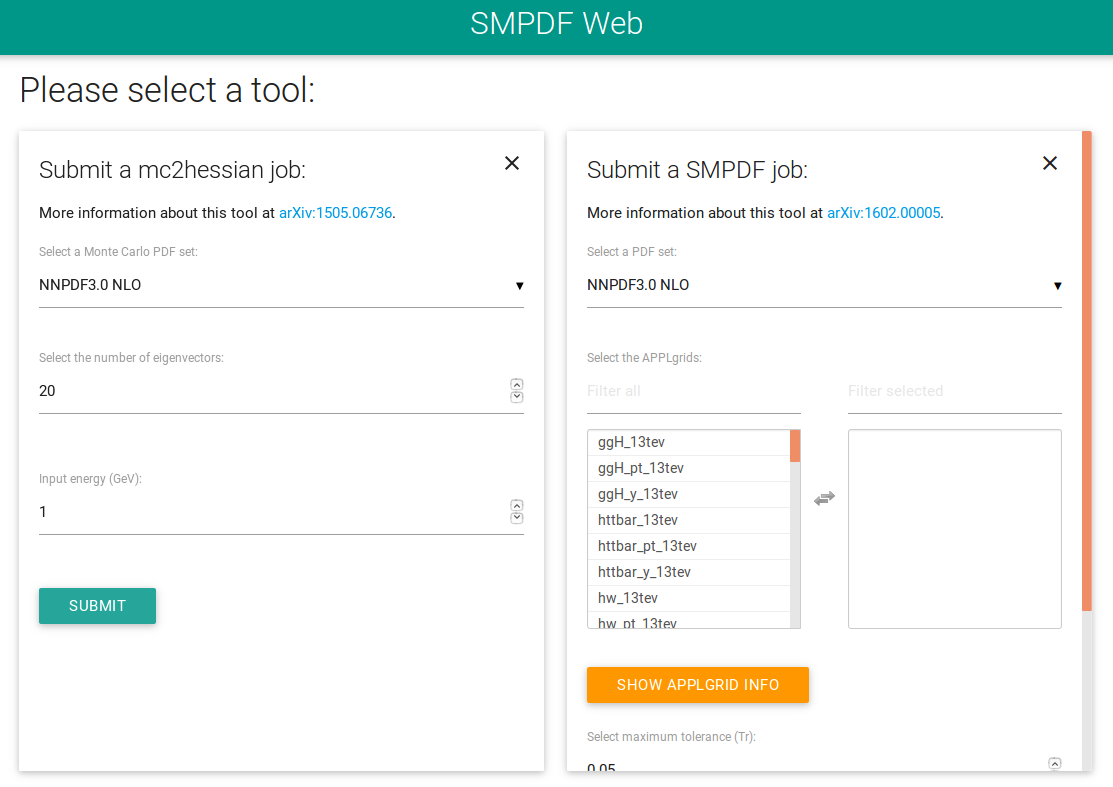}
    \par\end{centering}
  \caption{\label{fig:site} The SMPDF Web interface. The user
  can select the MC2H tool (left) or the SMPDF one (right)}
\end{figure}

In the outlook of the original paper we speculated about the 
integration of the SMPDF software in a web-based application such as
APFEL Web~\cite{Carrazza:2014gfa,Bertone:2013vaa,Bertone:2016lga}.
Here we describe such an application: SMPDF Web. It is
available at:
\begin{center}
\PoSspecialurl{http://smpdf.mi.infn.it/}
\end{center}

With this application one can generate customized SMPDF set, for the
desired prior PDF and observables, as well as a Hessian representation
of a given Monte Carlo PDF obtained with the MC2H algorithm. The
resulting PDF sets (in the LHAPDF6 format) can be downloaded from the
result page, together with complementary information about the
procedure (in particular the resulting linear transformations) and
validation plots.  We have computed a selection of parton level
observables to be used for the SMDPF algorithm.

SMPDF Web is based on the public SMPDF code (where also the MC2H
algorithm is implemented), available at:

\begin{center}
\PoSspecialurl{https://github.com/scarrazza/smpdf}
\end{center}

\begin{figure}[t]
  \begin{centering}
    \includegraphics[scale=0.25]{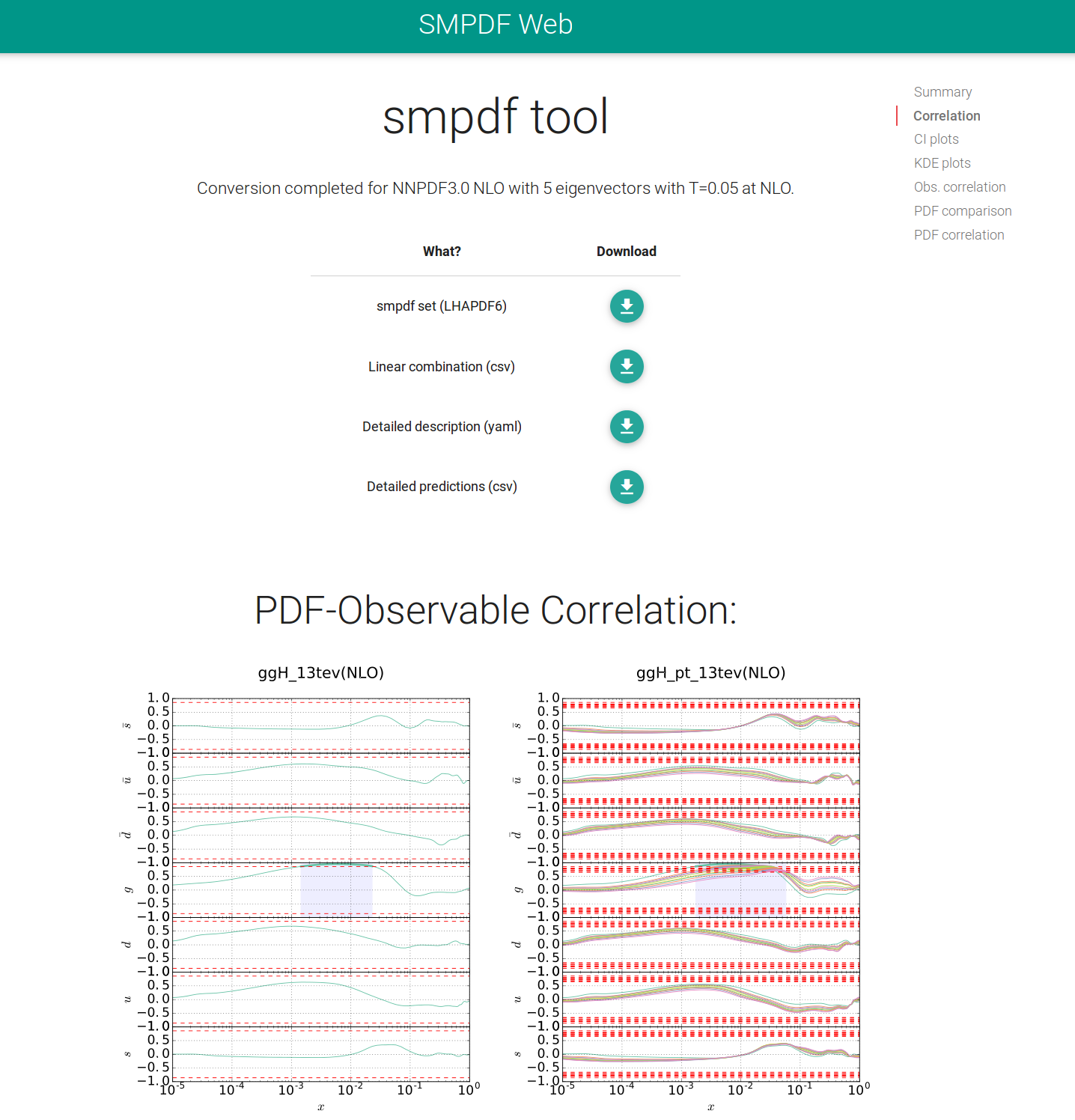}
    \par\end{centering}
  \caption{\label{fig:site2} The result page for the SMPDF tool. The
  user is presented with the option to download the resulting PDF set
  in the LHAPDF format, as well as other relevant outputs and
  validation plots (in the image, figures highlighting the kinematic
  PDF dependence of the $ggH$ process).}
\end{figure}

In Figure~\ref{fig:site} we show the front page of the website. The
user can select one of the two tools by clicking on the corresponding
thumbnail. Then a form is presented asking for the required
parameters. For the MC2H tool these are:
\begin{itemize}
\item The input PDF set, from a list of installed options.
\item The number of desired eigenvectors.
\item The energy scale at which the PDFs are evaluated, in GeV.
\end{itemize}
MC2H works best when the number of eigenvectors is high enough that
the whole PDF covariance matrix can be reproduced (a plot is shown as
part of the output).  In this case the energy scale makes little
impact as long as it is above the scale at which the evolution of the
input PDF can be considered reliable.

In turn, the SMPDF tool requires the following input:
\begin{itemize}
\item An input PDF set (MC or Hessian).
\item A list of observables.
\item The threshold parameter (tolerance), indicating the maximum
	allowable deviation in the reproduction of uncertainties.
\item The perturbative order (LO/NLO) at which the observables are computed.
\end{itemize}
The observables we provide in the web interface are in the APPLgrid
format~\cite{Carli:2010rw} and include those analyzed in
the SMPDF paper~\cite{Carrazza:2016htc}). The SMPDF code also allows
hadron level predictions entered directly in a text format. We find
that a tolerance of 5\% or 10\% results in a good compromise
between accuracy and the resulting number of error sets, as long as
the SMPDF set is used for observables compatible with those given as
input.

After the successful completion of the respective forms, a new page is
rendered, from which the user can download the resulting PDF set, as
well as other useful results (such as the linear transformations
involved in both algorithms), and showing several validation plots
(see Fig.~\ref{fig:site2}).
\begin{figure}[t]
  \begin{centering}
    \includegraphics[scale=0.3]{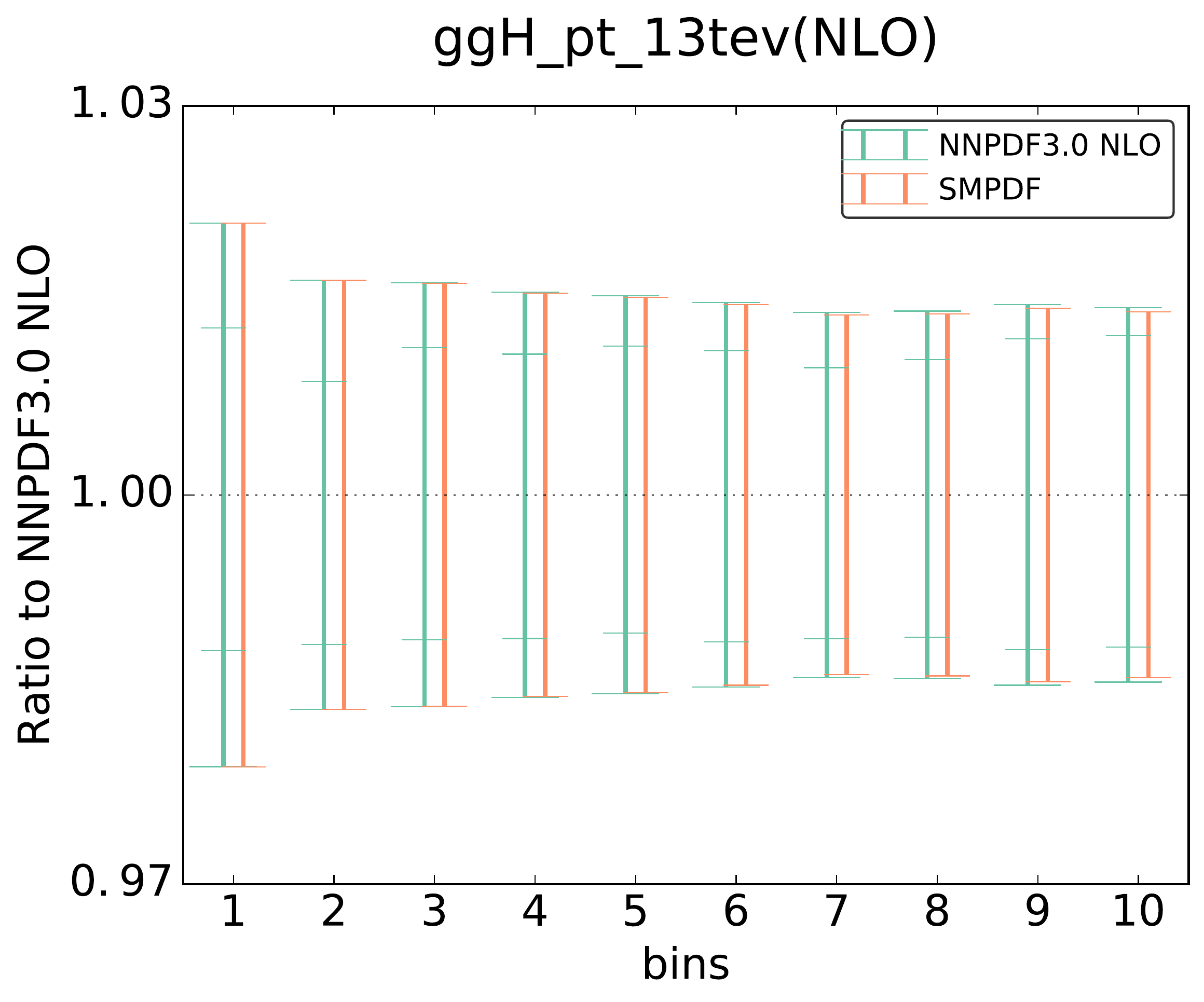}\includegraphics[scale=0.3]{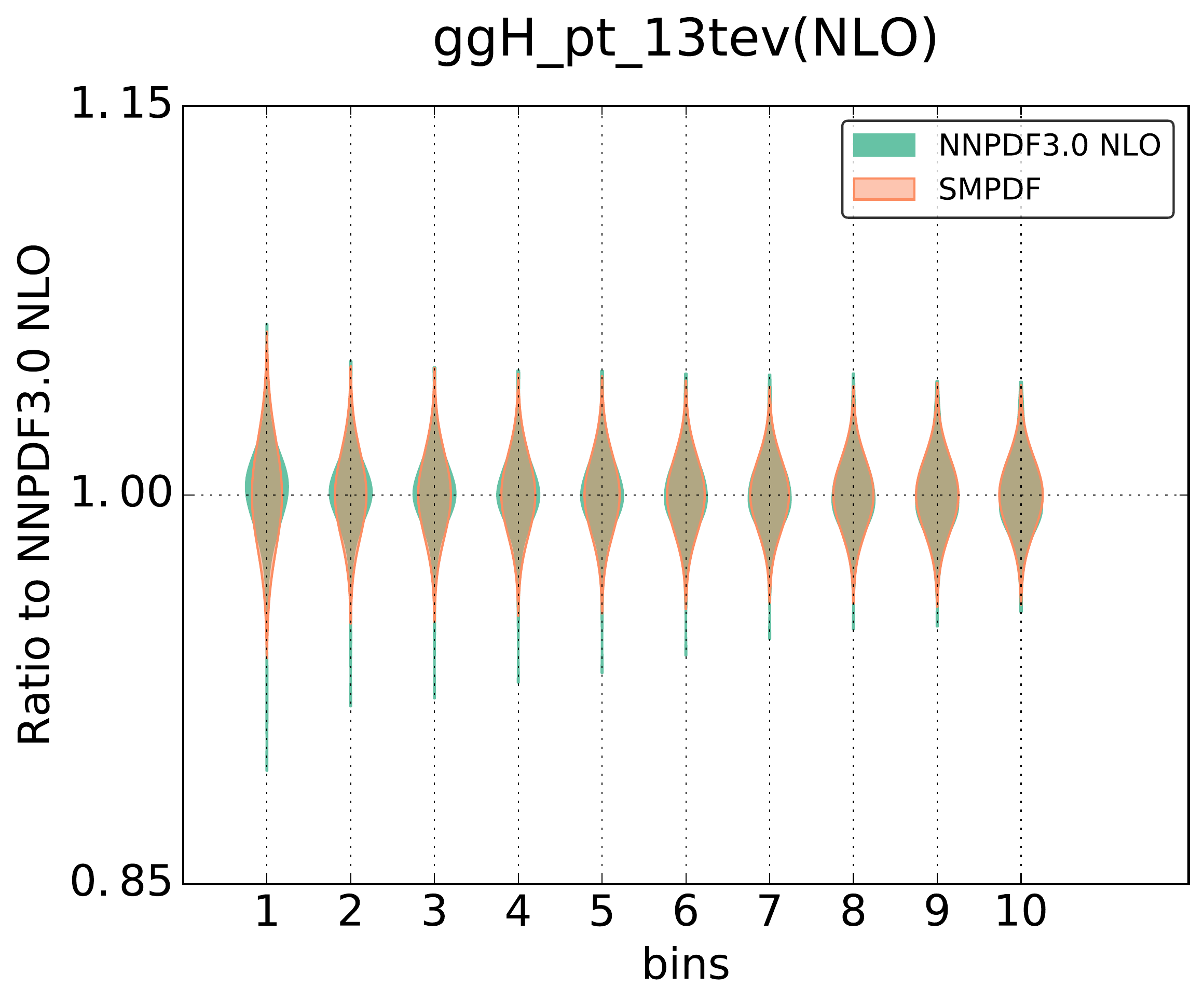}
    \par\end{centering}
  \caption{\label{fig:example1} Example of confidence interval (left)
    and kernel density estimate (right) predictions for the SMPDF for
    $ggH$ predictions.}
\end{figure}

\begin{figure}[t]
  \begin{centering}
    \includegraphics[scale=0.35]{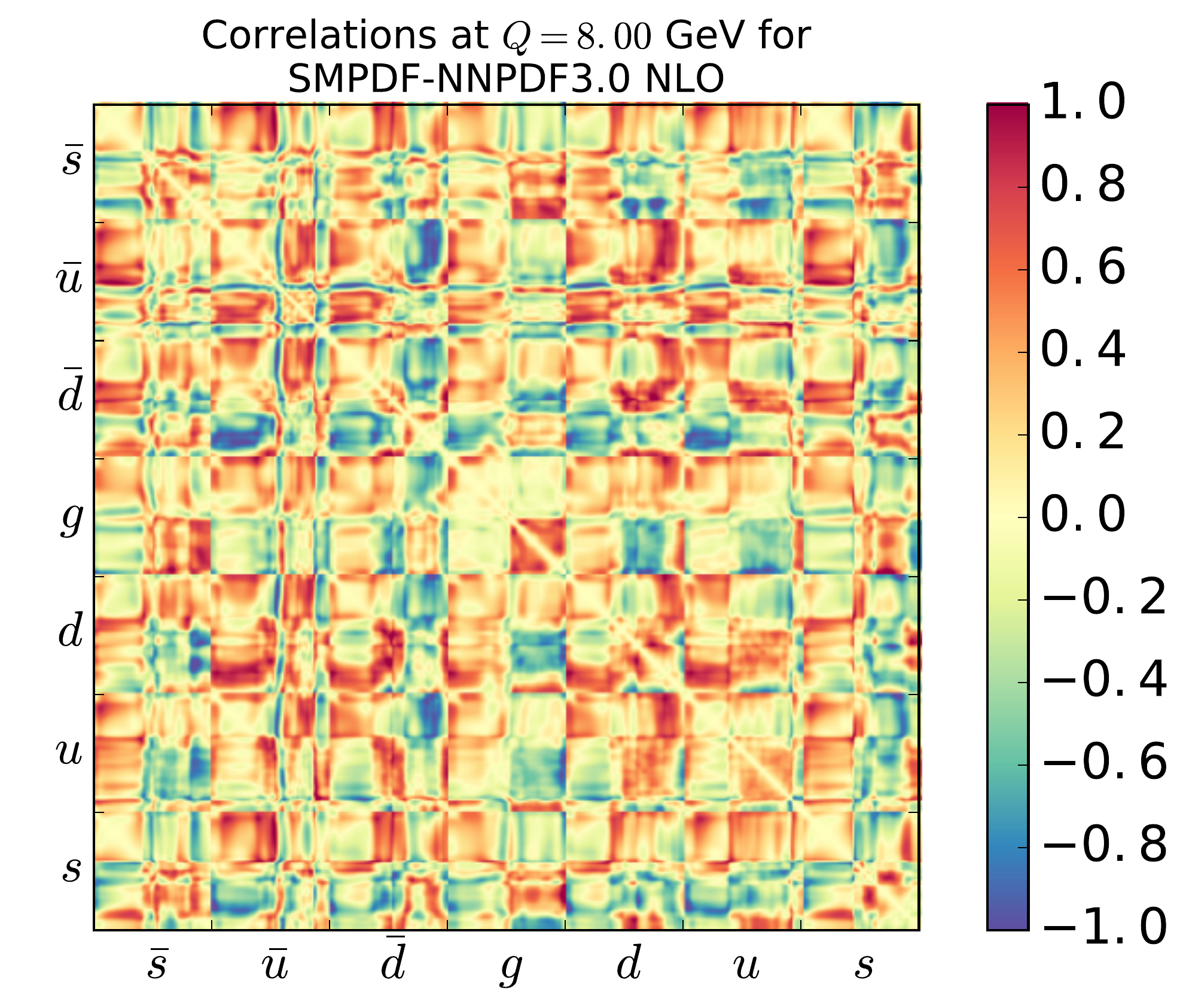}
    \par\end{centering}
  \caption{\label{fig:example2} Example of PDF-PDF correlations for
  SMPDFs optimized for Higgs production from gluon fusion. The most
  relevant correlations for this process (gluon at small $x$) are
  reproduced at percent level}
\end{figure}

Figures~\ref{fig:example1} and~\ref{fig:example2} show examples of
validation plots from SMPDF Web. Specifically we have generated
an SMPDF set using NNPDF3.0 NLO as a prior PDF and the $ggH$
$p_T$ distribution at NLO as input, with 5\% tolerance.
We obtain a Hessian set with $N_{\rm eig}=3$. Fig.~\ref{fig:example1}
left compares prediction uncertainties between the prior (green) and
the resulting SMPDF (orange), showing an almost perfect agreement,
and the right figure represents the same quantities as a 
kernel density estimate. Fig.~\ref{fig:example2} shows the PDF correlation
difference between the prior and the final SMPDF set. As expected, we
observe small differences for the gluon channel.

For the time being we provide a limited number of input PDF sets and
processes; however users are invited to submit upload requests
directly from the website.

\begin{figure}[t]
  \begin{centering}
    \includegraphics[width=0.7\textwidth]{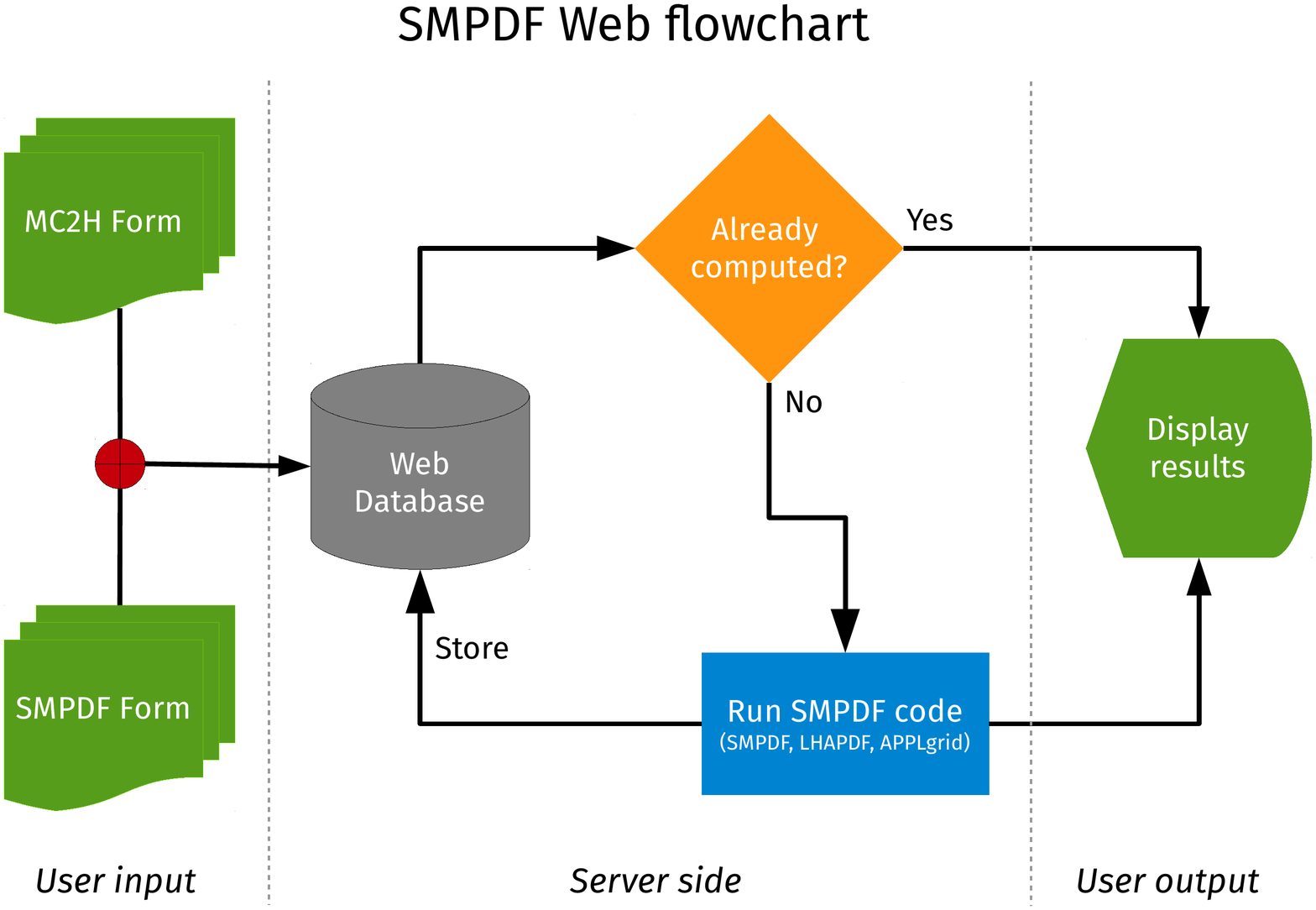}
    \par\end{centering}
    \caption{\label{fig:diagram} The SMPDF web application flowchart.}
\end{figure}

SMPDF Web implements a simple caching mechanism which stores
configuration and results. In this way we provide instantaneous
results for jobs which have been already computed, and the generated
URL can be used to reference and share the output. In
Fig.~\ref{fig:diagram} we show a diagram summarizing the layout of the
web application.

\paragraph{Acknowledgements}

S.C. is supported by the HICCUP ERC Consolidator grant (614577).
Z.K. is supported by  the  Executive  Research  Agency  (REA)  of  the
European Commission under the Grant Agreement PITN-GA-2012-316704
(HiggsTools).

\end{document}